\documentclass[a4paper,11pt]{article}
\usepackage{pos}

\usepackage{lineno}
\usepackage{amsmath}
\usepackage{dsfont} 
\usepackage{nicefrac}
\usepackage{braket}
\usepackage{enumitem}
\usepackage{mathtools}

\newcommand{\be}{\begin{equation}}
\newcommand{\ee}{\end{equation}}

\newcommand{\lint}{\big\langle}
\newcommand{\rint}{\big\rangle}

\newcommand{\sepc}{|}
\newcommand{\lec}{{_c\langle}}
\newcommand{\ric}{\rangle_c}

\DeclarePairedDelimiterX\mybraket[2]{\langle}{\rangle}{#1 \delimsize\vert #2}       
\DeclarePairedDelimiterX\mybraketOp[3]{\langle}{\rangle}%
{#1\,\delimsize\vert\,\mathopen{}#2\,\delimsize\vert\,\mathopen{}#3} 

\newcommand{\scprod}[2]{{{#1}\!\cdot\!{#2}}}

\newcommand{\Coll}[1]{C_{#1}}

\newcommand{\iden}{I}

\newcommand{\ISoft}{I_{S}}

\newcommand{\IVirt}{I_{V}}

\newcommand{\IColl}{I_{C}}

\newcommand{\ITot}{I_{T}}

\newcommand{\ICatbar}{\overline{I}_1}

\newcommand{\ColT}[1]{\boldsymbol{T}_{#1}}

\newcommand{\rmd}{\mathrm{d}}
\newcommand{\sigmahat}{\hat{\sigma}}
\newcommand{\obs}{\mathcal{O}}

\newcommand{\ww}[1]{\omega^{#1}}

\newcommand{\LO}{\mathrm{LO}}
\newcommand{\NLO}{\mathrm{NLO}}
\newcommand{\NNLO}{\mathrm {NNLO}}
\newcommand{\order}[1]{\mathcal{O}(#1)}
\newcommand{\ONLO}{\mathcal{O}_\text{NLO}}

\newcommand{\PAP}{\hat P^{(0)}}

\newcommand{\hatP}{\hat{P}}
\newcommand{\CalP}{\mathcal{P}}

\newcommand{\CalPgen}{\mathcal{P}^{\mathrm{gen}}}

\newcommand{\PNLO}{\mathcal{P}^{\mathrm{NLO}}}
\newcommand{\PaaNLO}{\PNLO_{aa}}

\newcommand{\CalPfin}{\mathcal{P}^{\mathrm{fin}}}

\newcommand{\amp}{\mathcal{M}}
\newcommand{\FLM}{F_{\mathrm{LM}}}

\newcommand{\FLV}{F_{\mathrm{LV}}}
\newcommand{\FLMcol}{\tilde{F}_{\mathrm{LM}}}
\newcommand{\FLRV}{F_{\rm RV}}
\newcommand{\FLVfin}{F_{\mathrm{LV}}^{\mathrm{fin}}}

\newcommand{\ampM}[1]{\mathcal{M}_{#1}}

\newcommand{\Emax}{E_{\rm max}}

\newcommand{\eps}{\epsilon}
\newcommand{\ep}{\epsilon}
\newcommand{\musq}{\mu^2}
\newcommand{\Np}{N_p}

\newcommand{\colsing}{X}

\newcommand{\qb}{\bar{q}}

\newcommand{\fl}[1]{f_{#1}}

\newcommand{\Ca}{C_A}

\newcommand{\asbr}{[\alpha_s]}
\newcommand{\amu}{\frac{\alpha_s(\mu^2)}{2\pi}}

\newcommand{\conv}{\otimes}

\newcommand{\hypF}{{_2F_1}}
\newcommand{\THmn}{\Theta_{\Fp \Sp}}   
\newcommand{\T}{\boldsymbol{T}}

\newcommand{\colorprod}{\cdot}

\newcommand{\Fp}{\mathfrak{m}}
\newcommand{\Sp}{\mathfrak{n}}

\newcommand{\oS}{\overline S}
\newcommand{\oC}{\overline C}

\title{Advances in the nested soft-collinear subtraction scheme}

\author*[a,b]{Chiara Signorile-Signorile}
\author[c]{Davide Maria Tagliabue}

\affiliation[a]{Institute for Institut f{\"u}r Theoretische Teilchenphysik, Karlsruher Institut f{\"u}r Technologie, Karlsruhe, Germany}

\affiliation[b]{Institut f{\"u}r Astroteilchenphysik, Karlsruher Institut f{\"u}r Technologie, D-76021 Karlsruhe, Germany}

\affiliation[c]{Tif Lab, Dipartimento di Fisica, Universit\`a di Milano and INFN, Sezione di Milano, Via Celoria 16, I-20133 Milano, Italy}

\emailAdd{chiara.signorile-signorile@kit.edu}
\emailAdd{davide.tagliabue@unimi.it}

\abstract{
We discuss a path~\cite{Devoto:XXXX} toward the generalisation of the nested soft-collinear
subtraction scheme~\cite{Caola:2017dug} to arbitrary $2\rightarrow n$ processes. The scheme is designed to provide an efficient and process-independent procedure to extract and regulate infrared (IR) singularities arising from unresolved real radiation and combine them with explicit singularities in virtual corrections. The new approach is based on a reorganisation of the relevant subtraction terms into simple combinations of a relatively small number of recurring  structures. This strategy leads to a drastic reduction in the computational effort required 
to derive integrated subtraction terms, while preserving the full generality of the scheme.  
We believe that this approach will allow for tackling the issue of regularising IR divergences at next-to-next-to-leading order in the strong coupling constant for arbitrary, multi-parton processes.

\bigskip

P3H-23-044, \, TTP23-025, \, TIF-UNIMI-2023-15
}

\FullConference{16th International Symposium on Radiative Corrections: Applications of Quantum Field Theory to Phenomenology (RADCOR2023)\\
28th May - 2nd June, 2023\\
Crieff, Scotland, UK\\}

\begin{document}
\maketitle

\section{The problem, the idea}
Data are expected to become more precise in the upcoming high-luminosity run of the Large Hadron Collider. The drastic reduction in experimental uncertainties expected in the
near future challenges the
theory community to provide more accurate theoretical predictions. In addition to multi-loop amplitudes, 
the relevant fixed-order calculations rely on the efficient treatment of infrared singularities. These singularities cancel out when combining virtual corrections, phase-space integrals of unresolved radiation, and collinear renormalisation of parton distribution functions (PDFs). 
While understanding of these singularities is well advanced (see Ref.~\cite{Agarwal:2021ais} for a recent review), developing optimal algorithms for infrared subtraction beyond next-to-leading order (NLO) remains an open issue. Various methods have been proposed~\cite{Frixione:2004is, Gehrmann-DeRidder:2005btv, Currie:2013vh,Somogyi:2005xz,Somogyi:2006db,Czakon:2010td,Czakon:2011ve,Anastasiou:2003gr, Caola:2017dug,Catani:2007vq,Grazzini:2017mhc, Boughezal:2011jf,Gaunt:2015pea,Sborlini:2016hat,Herzog:2018ily,Magnea:2018hab,Bertolotti:2022aih,Capatti:2019ypt} (for a recent review see Ref.~\cite{TorresBobadilla:2020ekr}), but achieving universality and efficiency at NNLO is still a challenge.  

In this proceeding we present a generalisation of
the nested soft-collinear subtraction scheme~\cite{Caola:2017dug} to parton annihilation into a final state with arbitrary number of gluons and colorless partons. 
The scheme was used to provide differential distributions for color singlet production~\cite{Caola:2019nzf} and decay~\cite{Caola:2019pfz}, deep inelastic scattering~\cite{Asteriadis:2019dte}, Higgs production in WBF~\cite{Asteriadis:2021gpd}, non-factorisable corrections to $t$-channel single-top production~\cite{Bronnum-Hansen:2022tmr}, as weel as mixed QCD-electroweak corrections to
$W$ and $Z$-boson production~\cite{Buccioni:2020cfi, Behring:2020cqi} and 
to neutral-current-mediated   production of a pair of massless leptons~\cite{Buccioni:2022kgy}. Existing applications share common features: 
given a small number of 
colored partons at leading order, the color algebra is trivial and the number of subtractions terms
is limited, providing a significant advantage for the bookkeeping. Moreover, the pole cancellation is performed analytically, after the explicit evaluation of each subtraction term separately. \\
This process-driven approach, although successful for simple processes, becomes
unfeasible for higher-multiplicity processes. 
To solve this problem, we suggest to preserve the full dependence on color degrees of freedom, and appropriately 
combine counterterms, prior to 
their explicit calculation. This strategy provides important insights 
into the general organisation of the subtraction procedure and into the roles of different terms, making the procedure more transparent.

At variance with previous applications, the present
study addresses multi-parton processes in full generality. The key idea relies on organising subtraction terms into recurring building blocks that combine to form finite quantities. This approach allows for many simplifications among different counterterms, before their evaluation. As a result, a remarkable improvement in simplicity and physical transparency in the construction of the subtraction terms is achieved. \\
In this
manuscript we provide a brief overview on some results that will be extensively discuss in Ref.~\cite{Devoto:XXXX}. In particular, we present a fully general formula for the NLO correction descriptive of generic $2 \rightarrow X+ n\,g$ processes. Additionally, we discuss a few points related to the application of this  novel approach to NNLO.

\section{Tackling the next-to-leading order correction}
In this section we discuss the calculation of QCD corrections to $p p \rightarrow  X + n \, g$ production at NLO, where $n$ is the number of final-state gluons and $X$ denotes a generic system of colorless particles. The differential cross section receives contributions from the virtual corrections, from the real emission of one additional gluon, and from the collinear renormalisation of PDFs
\be
\rmd \sigmahat^{\NLO} = \rmd \sigmahat^{\rm V}+ \rmd \sigmahat^{\rm R}+ \rmd \sigmahat^{\rm pdf} \, .
\ee
Our goal is to express each of the above contributions in terms of simple structures, with specific kinematic and color properties.
For a generic process, we
treat matrix elements 
as vectors in color space.
A matrix element where partons are assigned definite color indices is then written 
as a projection on a particular color-space basis vector. Denoting  the lowest-order matrix element for the partonic process $ab \to \colsing + n \, g$
as $\sepc \ampM{0}(1_a,2_b;... \, ,\Np ; \colsing) \ric \equiv \sepc\ampM{0}\ric$, $\Np= n+2$,
we introduce a function 
\be
\begin{split}
\widetilde{F}_{\rm LM}(1_a,2_b; ... \, ,  \Np; \colsing) =  \sepc \ampM{0} \ric \otimes \lec  \ampM{0} \sepc \,  {\rm dLips}_{\rm \colsing} \, \obs(\{p_i\}) \, ,
\label{eq:defnFLMcol}
\end{split}
\ee
where $\otimes$ indicates a tensor product in color space, ${\rm dLips}_{\rm \colsing}$ is the Lorentz-invariant phase space for the colorless system $\colsing$, and $\obs$ is an infrared-safe observable.  Taking the trace in color space gives the matrix element squared
$\mathrm{Tr} \big[ \widetilde{F}_{\rm LM}  \big] = {\rm dLips}_{\rm \colsing} \; 
| {\cal M} |^2 \; \obs \equiv \FLM$. 
When acting on $\FLMcol$ with a function $A$ of operators in color space, we define the notation
\be
\begin{split}
    A \colorprod \FLM &\equiv \mathrm{Tr} \big[ A \, \widetilde{F}_{\rm LM} \big]_c  = \lec \ampM{0} \sepc A \sepc \ampM{0} \ric \; {\rm dLips}_{\rm \colsing} \; \obs.
    \label{eq:colorproddefn}
    \end{split}
\ee
The LO partonic cross section is obtained by integrating $\FLM$ over the phase space of the final-state partons
\be 
\rmd \sigmahat^{\LO} = \int  \prod_{i=3}^{N_p} \; [\rmd p_i] \; \FLM(1_a,2_b; ... \, , \Np ;\colsing)
= \lint \FLM \rint \, , \qquad
 [\rmd p_i] = \frac{d^3 p_i}{(2\pi)^3 2 E_i} \, .
\label{eq:defnFLM1}
\ee
We then consider the virtual corrections, and examine the one-loop amplitude $\ampM{1}$. Infrared singularities of $\ampM{1}$ are given by the Catani's formula~\cite{Catani:1998bh}
\begin{align}
    \ampM{1}(1_a,2_b; ... \, \Np;X) = [\alpha_s] \; \ICatbar(\eps) \; \ampM{0}(1_a,2_b; ... \,  \Np;X)+ \ampM{1}^{\rm fin}(1_a,2_b; ... \,  \Np;X) \; , 
\label{eq:oneloop}
\end{align}
with 
\be
    \ICatbar(\eps) = \frac{1}{2}\sum_{(ij)} \frac{\mathcal{V}_i^\text{sing}(\eps)}{\ColT{i}^2}   \scprod{\ColT{i}}{\ColT{j}} \left(\frac{\musq} {2p_i\cdot p_j}\right)^\epsilon e^{i \pi\lambda_{ij} \ep} \; ,
    \qquad
    \mathcal{V}_i^\text{sing}(\eps) = \frac{\ColT{i}^2}{\epsilon^2} + \frac{\gamma_i}{\epsilon} \; .
 \label{eq:I1Cat}
\ee
In Eq.~\eqref{eq:I1Cat}
the sum runs over all pairs of distinct partons; definitions of all the relevant quantities can be found in Ref.~\cite{Catani:1998bh}. Here we just mention that $\ampM{1}^{\rm fin}$ is the infrared-finite remainder of the one-loop amplitude, $[\alpha_s] = \amu (e^{\eps \gamma_\text{E}})/\Gamma(1-\eps)$, where $\alpha_s(\mu^2)$ is the  renormalised strong coupling constant, and $\ColT{i}$ is the color-charge operator of parton $i$. Squaring the amplitude we find the virtual correction to the cross section
\be
    \rmd \sigmahat^V = 
    \lint \FLV \rint = 
    [\alpha_s] \lint \IVirt(\eps) \colorprod \FLM \rint +  \lint \FLVfin \rint \; ,
    \label{eq:I1Virt_defn}
\qquad 
    \IVirt(\eps) = \ICatbar(\ep) + \ICatbar^\dagger(\ep) \, .
\end{equation}
Here $\FLM$ represents the LO matrix element squared, and the angular brackets imply the integration over the fiducial final-state phase space. Normalisation factors accounting for color and spin averages, as well as symmetry factors, are included in the definition of $\FLM$. We emphasise that the functions describing $\rmd \sigmahat^V$ are naturally affected by color correlations, i.e.~by terms proportional to $\scprod{\ColT{i}}{\ColT{j}}$, as a consequence of the definition of $\ICatbar$. \\
We then turn to the real-radiation contribution $\rmd \sigmahat^{\rm R} =   \lint  \FLM(\Np+ 1;\colsing) \rint$. In order to compute it, we need to extract the singularities that appear when \emph{any} of the $(n+1)$ final state gluons becomes either soft or collinear to another parton. To isolate potentially unresolved gluon, we introduce the damping factors $\Delta^{(i)}$, which add up to 1 and ensure that the only unresolved parton in $\Delta^{(i)} \FLM$, with $i=1, ..., \Np+1$, is the parton $i$.
Given that $\FLM$ is unchanged under any permutation of the final state gluons, we extract the relevant symmetry factor from $\FLM$ and obtain
\be
\label{eq:damping}
\frac{1}{(n+1)!} 
\sum_{i=3}^{\Np+1} \lint\Delta^{(i)} \FLM(1_a,2_b; ... \, \Np + 1;\colsing) \rint 
 = \lint \Delta^{(\Fp)}  \FLM(\Fp) \rint \; .
\ee
In Eq.~\eqref{eq:damping} the gluon $\Fp_g$ can potentially become unresolved, while the remaining $\Np$ partons are resolved.
We note that in Eq.~\eqref{eq:damping}
the symmetry factor contributing to the r.h.s.~is understood to be $1/n!$ and it is reabsorbed in the definition of $\lint \Delta^{(\Fp)}  \FLM(\Fp) \rint$.
 We now proceed to subtract the singularities in $\Delta^{(i)} \FLM$, starting with the soft one. We thus write 
\be
    \lint \Delta^{(\Fp)} \FLM(\Fp) \rint =
    \lint S_\Fp \FLM(\Fp) \rint
    + \lint \oS_\Fp \Delta^{(\Fp)} \FLM(\Fp ) \rint \; ,
\label{eq:softlimNLO}
\ee
where $S_\Fp$ is an operator
that extracts the leading soft
behaviour of $\Fp_g$ and acts on everything that is on its right\footnote{In Eq.~\eqref{eq:softlimNLO} we used $S_\Fp \Delta^{(\Fp)} = 1$.}, while $\oS_\Fp \equiv \iden - S_\Fp$.
The first term in Eq.~\eqref{eq:softlimNLO} corresponds to the soft limit of the squared matrix element, integrated over the unresolved phase space. It reads~\cite{Caola:2017dug,Devoto:XXXX}
\begin{align}
    \lint S_{\Fp}  \FLM( \Fp ) \rint =  
    - [\alpha_s] \frac{(2\Emax/\mu)^{-2\eps}}{\ep^2}  \sum_{(ij)} 
    \lint \eta_{ij}^{-\eps} K_{ij} \left(\scprod{\ColT{i}}{\ColT{j}} \right)\colorprod \FLM \rint
    \equiv [\alpha_s] \lint \ISoft (\eps)\colorprod \FLM \rint \; ,
\label{eq:NLO_soft}
\end{align}
where $\eta_{ij}=(1-\cos \theta_{ij})/2$, $K_{ij}\propto\eta_{ij}^{1+\epsilon} \hypF(1,1,1-\eps,1-\eta_{ij})$, and $\Emax$ can be chosen such that it exceeds the maximal energy of any parton in the considered process. We notice that, similar to the Catani's operator $\ICatbar$, also the soft operator $\ISoft$ in Eq.~\eqref{eq:NLO_soft} contains color-correlated contributions. \\
We now consider the second term on the right-hand side of Eq.~\eqref{eq:softlimNLO}. This term is soft-regulated but it is still affected by collinear singularities. In order to isolate them, we introduce partition functions $\ww{\Fp i}$, which satisfy the conditions $\Coll{j \Fp} \ww{\Fp  i} = \delta^{ij}$, $\sum_i \ww{\Fp i}=1$. 
Here $C_{ij}$ is the operator that extracts the leading behaviour of the matrix element under the limit $i\parallel j$. We are then able to write
\begin{align}
    \lint \oS_\Fp \Delta^{(\Fp)} \FLM(\Fp ) \rint
    =
    \sum_{i=1}^{\Np} 
    \lint \oS_\Fp C_{i \Fp } \Delta^{(\Fp)} \FLM(\Fp)  
    \rint
    + 
    \sum_{i=1}^{\Np} 
    \lint \oS_\Fp \oC_{i \Fp} \, \ww{\Fp i}   \Delta^{(\Fp)} \FLM(\Fp) \rint \; ,
\label{eq:NLOregulated}
\end{align}
with $\oC_{i\Fp} \equiv \iden - C_{i\Fp}$. 
The last term on the r.h.s.~is fully regulated and can be integrated numerically in four dimensions. The hard-collinear limits in the first term factorise into universal collinear functions and leading-order $\FLM$. We first consider the configurations where a final state gluon $[i\Fp]_g$ splits into two collinear gluons $\Fp_g$ and $i_g$. Integrating over the phase space of gluon $\Fp_g$, and summing over the helicities of $\Fp_g$ and $i_g$, for $i=3, ...\, , \Np$ we find
\be
\begin{split}
\label{eq:Coll}
  & \lint \oS_\Fp C_{i \Fp }\Delta^{(\Fp)} \FLM(\Fp)  \rint 
   = 
   [\alpha_s] \frac{\Gamma_{i,g}}{\eps} \lint \FLM\rint \, ,
\end{split}
\ee
where
\be
\begin{split}
    &
    \Gamma_{i,g} = \left(\frac{2E_i}{\mu}\right)^{-2\eps} \frac{\Gamma^2(1-\eps)}{\Gamma(1-2\eps)} \; \gamma_{z,g}^{22}   \; ,  
    \qquad
    \gamma_{z,g}^{22} = - \int_{0}^{1} \rmd z\, (1-S_z) \, 
    \frac{z \, \hatP_{gg}(z)}{z^{2 \eps} (1-z)^{2\eps}}  
    + \Ca \frac{1 - e^{2\eps L_i}}{\eps}  \; .
    \nonumber
\end{split}
\ee
In the above $z \equiv E_i/(E_\Fp+E_i)$, $S_z$ is the soft $z \rightarrow 1$ limit of the integrand, $L_i = \log(\Emax/E_i)$ and $\hat{P}_{gg}(z)$ is the gluon spin-averaged splitting function.
When the gluon $\Fp_g$ becomes collinear to the initial state parton $1_a$, $\FLM$ becomes dependent on the energy fraction $z=1-E_{\Fp}/E_a$. Therefore the integration over the energy of the gluon $\Fp_g$ results in the convolution of the appropriate splitting function and the boosted matrix element. Such convolution is identified with the symbol $\otimes$. Integrating over the angle between partons $\Fp_g$ and $a$, we obtain
\be
   \lint \oS_\Fp  C_{a\Fp}  \Delta^{(\Fp)} \FLM(\Fp ) \rint  =  [\alpha_s] \frac{\Gamma_{a,f_a}}{\epsilon} \lint \FLM\rint  +  
   \frac{[\alpha_s]}{\epsilon} \lint \CalPgen_{aa} \conv \FLM \rint \; .
   \label{eq:NLO_hard_coll_final}
\ee
Here $\Gamma_{a,f_a}$ is the \emph{generalized initial state anomalous dimension} ($f_a$ stands for the ``flavour of particle $a$''), i.e.
\begin{equation}
    \Gamma_{a,f_a} = \left(\frac{2E_a}{\mu}\right)^{-2\eps} \frac{\Gamma^2(1-\eps)}{\Gamma(1-2\eps)} \left( \gamma_{f_a} + \ColT{f_a}^2 \frac{1-e^{-2\eps L_a}}{\epsilon}  \right) \; , 
    \label{eq:gamma_expansion_is}
\end{equation}
where $\ColT{q}^2 =\ColT{\bar q}^2= C_F$, $\ColT{g}^2 = C_A$. In Eq.~\eqref{eq:NLO_hard_coll_final} we also introduced   the \emph{generalized splitting function} $\CalPgen_{aa}$. It reads 
\begin{align}
\CalP_{\fl{a} \fl{a}}^{\text{gen}}(z) = & \left(\frac{2E_a}{\mu}\right)^{-2\epsilon} \frac{\Gamma^2(1-\epsilon)}{\Gamma(1-2\epsilon)} \left[- \PAP_{\fl{a} \fl{a}}(z) + \epsilon \,  \CalP_{\fl{a} \fl{a}}^{(k),\text{fin}}(z)\right] \; ,
\label{Eq:Paa_GEN_definition_k_general}
\end{align}
with $\PAP_{\fl{a} \fl{a}}(z)$ being the Altarelli-Parisi splitting function in four dimensions, and $\CalP_{\fl{a} \fl{a}}^{(k),\text{fin}}(z)$ the $\order{\ep}$ remainder. 
Eqs.~\eqref{eq:Coll} and ~\eqref{eq:NLO_hard_coll_final} can be split according to the kinematic dependence of $\FLM$ into a boosted contribution, which depends on $\FLM(z)$, and 
an elastic part, proportional to $\FLM=\FLM(z=1)$. The latter can be organised into a single-collinear "operator" 
\be
    \IColl(\eps) = \sum_{i=1}^{\Np}  \frac{\Gamma_{i,f_i}}{\epsilon} \; ,
\label{eq:IColl_definition}
\ee
acting on $\FLM$. We note that $\IColl(\eps)$ is singular starting at $1/\ep$.
Before combining all the integrated subtraction terms, we stress that 
collinear limits are local in the color space. For this reason, they do not contain color correlations and they are  proportional to Casimir operators only. The sum of all contributions to the real-emission cross section reads\footnote{Notice that the initial state of the process $ab \to X + n\, g$ can only be $q\qb$ or $gg$, which implies $\CalPgen_{bb}(z) \equiv \CalPgen_{aa}(z)$. We use the convention that $\lint\CalPgen_{aa}(z) \conv \FLM\rint$ indicates a convolution on the first leg and $\lint \FLM \conv \CalPgen_{aa}(z) \rint$ on the second.}
\be
\begin{split}
    \rmd \sigmahat^R ={}&
    [\alpha_s]
    \lint \big[
    \ISoft (\eps)
    +
    \IColl(\eps)
    \big]
    \colorprod \FLM \rint 
    +\frac{[\alpha_s]}{\eps} 
    \Big[\lint \CalPgen_{aa} \conv \FLM\rint 
    + \lint \FLM \conv \CalPgen_{aa} \rint\Big]  \\
    & + \sum_{i=1}^{\Np} 
    \lint \oS_\Fp \oC_{i \Fp} \, \ww{\Fp i} \Delta^{(\Fp)} \FLM(\Fp) \rint \; .
    \label{eq:NLO_real_poles}
\end{split}
\ee
Combining the virtual loop contribution in Eq.~\eqref{eq:I1Virt_defn} with Eq.~\eqref{eq:NLO_real_poles}, and focusing on the elastic piece, we identify  an infrared-finite quantity given by the sum of virtual, soft and single-collinear 
operators acting on $\FLM$
\be
\label{eq:IT_def}
    \lint \ITot(\eps) \colorprod \FLM \rint 
    =  
    \lint \big[\IVirt(\eps) + \ISoft(\eps) + \IColl(\eps)\big] \colorprod  \FLM\rint = \order{\eps^0} \; .
\ee
One can check that in the above expression all the singular color-correlated contributions vanish. This can be proven by noticing that the most  divergent terms in $\ICatbar$ and $\ISoft$ do not depend on the kinematics and therefore we can use color conservation to write 
\be
    \sum_{j \ne i}
    \mybraketOp*{\amp_0}{ \scprod{\ColT{i}}{\ColT{j}}
    }{\amp_0}  
    =  - \ColT{i}^2 \, |\ampM{0}|^2 \; .
    \label{eq:remove_col_corr}
\ee
The double poles then disappear due to the 
overall opposite sign of $\ICatbar$ with respect to $\ISoft$. The single pole coming from the virtual correction is now proportional to the logarithms of the Lorentz invariants $s_{ij}= 2p_i \cdot p_j$ and the color structure $\scprod{\ColT{i}}{\ColT{j}}$. We can decompose these logarithms into the sum of logarithms of energies and of angular variables. The latter are cancelled by corresponding terms arising from the soft counterterm. 
Terms containing logarithms of energies are manifestly dependent on one color index, say $i$, and therefore we can sum over the second index, say $j$, and use again the relation in Eq.~\eqref{eq:remove_col_corr}. We are then left with single poles multiplied by various constant and Casimir operators. These poles are exactly cancelled by the series expansion of $\IColl$, which is free of color correlations. The remaining poles in Eq.~\eqref{eq:NLO_real_poles} involve convolutions with $\CalPgen_{aa (bb)}$ and vanish against the collinear renormalisation of the PDFs.

Putting everything together, we are able to quote the finite expression for the NLO corrections to the LO process $1_a + 2_b \to X + n \, g$,
\begin{equation}
\begin{split}
    \rmd \sigmahat^{\NLO} = \rmd \sigmahat^{\rm V}+ \rmd \sigmahat^{\rm R}+ \rmd \sigmahat^{\rm pdf} 
    = & ~ [\alpha_s] \Big(
    \lint \ITot(\ep) \colorprod \FLM \rint 
    + \lint \PaaNLO \conv \FLM \rint + \lint \FLM \conv \PaaNLO \rint
    \Big)\\
    &  + \lint \ONLO \, \Delta^{(\Fp)} \FLM(\Fp) \rint + \lint\FLV^\text{fin}\rint \; .
    \label{Eq:final_expression_NLO}
\end{split}
\end{equation}
We have defined the subtraction operator for the fully-regulated contribution $\ONLO = \sum_{i} \oS_\Fp \oC_{i \Fp } \, \omega^{\Fp i}$, 
and introduced 
$\PaaNLO(z) = 2\log(2E_i/\mu)\PAP_{aa}(z)+ \CalPfin_{aa}(z)$.

\section{Comments on the NNLO calculation}
We now consider the NNLO corrections to $p p \rightarrow  X + n \, g$. Here we have contributions from the double-virtual, the real-virtual and the double-real corrections, as well as from the PDFs renormalisation, i.e.
\be
\label{eq:NNLO_sigma}
\rmd \sigmahat^{\NNLO} = \rmd \sigmahat^{\rm VV} + \rmd \sigmahat^{\rm RV} + \rmd \sigmahat^{\rm RR} + \rmd \sigmahat^{\rm pdf}.
\ee 
A systematic discussion of how to organise the NNLO calculation for a generic process is beyond the scope of the present manuscript, and we postpone it to a forthcoming paper \cite{Devoto:XXXX}. In this section we focus on a specific aspect of the NNLO subtraction procedure, namely the \emph{cancellation of double-color correlated terms}. Consider for instance the double-virtual component $\rmd \sigmahat^{\rm VV}$ of Eq.~\eqref{eq:NNLO_sigma}. Using the results of Ref.~\cite{Catani:1998bh}, one observes that $\rmd \sigmahat^{\rm VV}$ depends quadratically on the operator $\ICatbar$ defined in Eq.~\eqref{eq:I1Cat} and on the corresponding $\ICatbar^\dagger$. 
Since $\ICatbar \sim \T_i \cdot \T_j$, 
this means that $\rmd \sigmahat^{\rm VV}$ 
must contain terms of the type $(\T_i \cdot \T_j) (\T_k \cdot \T_l)$, 
which we define as {\it double-color correlated}. Analogous terms arise in both the double-soft contribution contained in $\rmd \sigmahat^{\rm RR}$ and in the single-soft term contributing to $\rmd \sigmahat^{\rm RV}$. Dealing with such double-color correlated terms is in general a non-trivial but well-defined problem. To solve it, we follow the same strategy as presented in the previous section on NLO corrections. In particular, we first isolate contributions in $\rmd \sigmahat^{\rm VV}$ that are affected by double-color correlations, and then we combine them with those contained within $\rmd \sigmahat^{\rm RR}$ and $\rmd \sigmahat^{\rm RV}$. The ultimate goal is to assemble all these double-color correlated terms into an expression that can be expresses as $\ITot^2$.
Once this is achieved, we are able to directly state that such correlations do not affect the pole content of Eq.~\eqref{eq:NNLO_sigma}, avoiding the need to evaluate them explicitly. 


With this goal in mind, we begin with the double-virtual correction. For the purposes of our discussion we restrict our analysis to the part of $\rmd \sigmahat^{\rm VV}$ that contains double-color correlations, i.e.\footnote{In order to simplify the notation, in the following we do not show dependencies of operators on $\ep$, unless these dependencies become important for the discussion.} 
\begin{align}
    Y_{\rm VV} =  \frac{\asbr}{2} 
    \mybraketOp*{\amp_0}{\ICatbar^{\, 2} 
    +  \big(\ICatbar^\dagger\big)^2 
    + 2 \ICatbar^\dagger 
    \ICatbar}{\amp_0} \; .
    \label{Eq:Xvv_def}
\end{align}
Notice that the combination in Eq.~\eqref{Eq:Xvv_def} is quite close to $\big(\ICatbar+\ICatbar^\dag\big)^2$. This observation suggests that, since we want to rewrite $Y_{\rm VV}$ in terms of $\ITot$, the latter must appear in second power, as already anticipated above. As said, we need other contributions of the type $\ISoft^2$, $\ISoft \IColl$, $\IColl^2$, $\ISoft \ICatbar^{(\dagger)}$ and $\ICatbar^{(\dagger)} \IColl$ in order to reconstruct $\ITot^2$. We need to find such structures among those arising from the double-real and the real-virtual corrections. \\
To do so, we first consider the double-real contribution $\rmd \sigmahat^{\rm RR}$. Following the iterative procedure described in Ref.~\cite{Caola:2017dug}, we extract the double-soft singularities that correspond to singular limits of unresolved partons $\Fp_g$ and $\Sp_g$. We find
\begin{align}
    \rmd \sigmahat^{\rm RR} = & ~ \lint S_{\Fp \Sp} \Delta^{(\Fp \Sp)} \THmn \FLM(\Fp,\Sp)\rint 
    + \lint \oS_{\Fp \Sp} S_\Sp \, \Delta^{(\Fp \Sp)} \THmn \FLM(\Fp,\Sp)\rint 
    \nonumber \\
    & + \lint \oS_{\Fp \Sp} \oS_\Sp \, \Delta^{(\Fp \Sp)} \THmn \FLM(\Fp,\Sp)\rint \; ,
\label{eq:double-real_subtra}
\end{align}
where $\THmn = \Theta(E_\Fp - E_\Sp)$ is the energy ordering that, together with the partition $\Delta^{(\Fp \Sp)}$ (a straightforward generalisation of $\Delta^{(\Fp)}$ to two unresolved partons), reabsorbs the symmetry factor of the matrix element. 
The first term in 
Eq.~\eqref{eq:double-real_subtra} is the 
double-soft unresolved terms. It contains
two main contributions: a single color-correlated and a double color-correlated component. We are interested in the second one, which is proportional to the factorised product of two eikonal functions and a LO matrix element. Upon integrating over the unresolved phase space, such term becomes 
\begin{equation}
    Y_{\rm RR}^{\rm (ss)}= \frac12 \, \mybraketOp*{\amp_0} { \ISoft^{\, 2} (\eps) }{\amp_0} \; .
    \label{Eq:XRR_ss_def}
\end{equation}
The second term in
Eq.~\eqref{eq:double-real_subtra} is the single-soft subtraction term. It is free of double-soft singularities, but still contains collinear configurations that need to be regularized as well. As expected, after some manipulations we can extract from it the simple combination of soft and collinear operators 
\begin{align}
\label{eq:SHCRR}
Y_{\rm RR}^{\rm (shc)}=
     & ~ \mybraketOp*{\amp_0}{\ISoft \, \IColl}{\amp_0} \; .    
\end{align}
The last term in 
Eq.~\eqref{eq:double-real_subtra} is fully soft-regulated, but still affected by collinear singularities that are 
isolated by a phase-space partitioning inspired by the FKS scheme~\cite{Frixione:1995ms}. Among the different terms that contribute to the soft-regulated piece, we are interested in the double-collinear component, which is proportional to two collinear limits. After some manipulations we identify the right piece
\begin{align}
\label{eq:CC}
    Y_{\rm RR}^{\rm (cc)} = \frac12 \,  \mybraketOp*{\amp_0}{\IColl^2}{\amp_0} \; . 
\end{align}
At this point the only missing contributions are of type $\ISoft \ICatbar^{(\dagger)}$ and $\ICatbar^{(\dagger)} \IColl$. To find where they come from, we have to consider the real-virtual corrections, which can be treated in full
analogy to the NLO case. In this regard, $\rmd \sigmahat^{\rm RV}$ can be written as 
\begin{equation}
\label{eq:RV}
    \rmd \sigmahat^{\rm RV} \equiv
    \lint \Delta^{(\Fp)} \FLRV(\Fp) \rint 
    = \lint S_\Fp \FLRV(\Fp) \rint 
    + \sum_{i=1}^{\Np} 
    \lint \oS_\Fp C_{i\Fp} \Delta^{(\Fp)} \FLRV(\Fp) \rint 
    + \lint \ONLO \, \Delta^{(\Fp)} \FLRV(\Fp) \rint \, . 
\end{equation}
The first term on the r.h.s.~describes kinematic configurations where parton $\Fp_g$ becomes soft. Among the different parts 
contributing to this term, we focus
on the component proportional to the eikonal function multiplied by the one-loop,  color-correlated matrix element. Upon integration over the soft-radiation phase space, we find this term to be equal to
\begin{align}
    Y_{\rm RV}^{\rm (s)} = [\alpha_s]^2 \mybraketOp*{\amp_0}{\ISoft \, \ICatbar + \ICatbar^\dagger \, \ISoft}{\amp_0} \, .
    \label{Eq:XRV_s_def}
\end{align}
Finally, the second term in Eq.~\eqref{eq:RV}
is the hard-collinear limit of RV. Again, we select only the factorised contribution and rewrite it in analogy to the soft component. We find
\begin{align}
\label{eq:HCRV}
    Y_{\rm RV}^{\rm (hc)} = \mybraketOp*{\amp_0}{\big(\ICatbar + \ICatbar^\dagger\big) \IColl}{\amp_0} \; .
\end{align}
We are now in the position to collect all the ingredients we were looking for. We combine $Y_{\rm VV}$, $Y_{\rm RR}^{\rm (ss)}$, $Y_{\rm RR}^{\rm (shc)}$, $Y_{\rm RR}^{\rm (cc)}$, $Y_{\rm RV}^{\rm (s)}$ and $Y_{\rm RV}^{\rm (hc)}$ to get an object $Y$ such that\footnote{The combination of these objects also contains commutators, since $\ICatbar$, $\ICatbar^\dagger$ and $\ISoft$ do not commute with each other. They give rise to terms proportional to $ f_{abc} \, T_i^a T_j^b T_k^c$, which play a different role. We refer this discussions to Ref.~\cite{Devoto:XXXX}.}
\begin{equation}
\begin{split}
    Y = \frac{\asbr^2}2 \mybraketOp*{\amp_0}{ \big[\IVirt + \ISoft + \IColl\big]^2}{\amp_0}
    \equiv
    \frac{\asbr^2}2 \mybraketOp*{\amp_0}{ \ITot^{\, 2}}{\amp_0} \; .
\end{split}
\end{equation}
This expression is equivalent to the square of the color-correlated term encountered at  NLO. 

The discussion illustrates the point that it is beneficial to combine certain subtracted term before attempting to evaluate them. This is particularly important for color correlated contributions, where iterative structures similar to those encountered at NLO appear.
As a further example, let us consider again $Y_{\rm VV}$ in Eq.~\eqref{Eq:Xvv_def}. 
Clearly this quantity contains poles starting from $\order{\ep^{-4}}$ and color correlations from $\order{\ep^{-3}}$. The same holds for $Y_{\rm RR}^{\rm (ss)}$ in Eq.~\eqref{Eq:XRR_ss_def} and $Y_{\rm RV}^{\rm (s)}$ in Eq.~\eqref{Eq:XRV_s_def}. In principle, nothing prevents us from calculating each of these objects individually and then summing them together (as was done in Ref.~\cite{Caola:2017dug}). 
As long as $\Np \le 3$, there are no conceptual difficulties in doing so, since the products $\T_i \cdot \T_j$ can always be written as combinations of Casimir operators. 
In contrast, if $\Np \ge 4$, the products $\T_i \cdot \T_j$ are matrices in color space. 
Using the approach of Ref.~\cite{Caola:2017dug}, one would have to deal with cancellations of poles that contain color correlations coming from a plethora of different terms already at $\order{\ep^{-3}}$. 
If one also requires the parameter $\Np$ to be generic, the problem of color correlations for a generic process would immediately become non-trivial. 
For these reasons, the method proposed in this paper changes the philosophy behind the organisation of infra-red subtraction. 
All the contributions are written in terms of universal operators as $\IVirt$, $\ISoft$, $\IColl$, which are then combined into IR-finite sub-blocks, as $\ITot$, in a way that it is possible to cancel double-color correlated poles without having to compute them explicitly first. 
This, in conclusion, leads to two important advantages: the problem of color correlations is solved, since everything is reabsorbed within $\ITot$, and the final result applies to a general final state.

\section{Conclusions}
We have discussed the recent progress in applications of the nested soft-collinear subtraction scheme~\cite{Caola:2017dug} for generic hadron collider process.
We have reduced the singular limits contributing to the NLO cross section to simple, recurring structures, which combine into compact, $\epsilon$-finite expression. This procedure has been discussed in details at NLO for gluon final states, but can be easily extended to accommodate final state quarks. At NNLO, the structures identified at NLO have been extracted to prove the cancellation of singular double-color-correlated contributions. This result points towards an opportunity of systematically combining NNLO subtraction terms into finite contributions, while minimising the need for explicit calculations. We believe our results to be a significant step toward the 
complete generalisation of the  subtraction scheme introduced in Ref.~\cite{Caola:2017dug} to account for an arbitrary number of final-state particles at NNLO. We will present 
further details of this study in an upcoming publication~\cite{Devoto:XXXX}.

\section*{Acknowledgments}

We thank Federica Devoto, Kirill Melnikov and  Raoul R\"ontsch for the collaboration on the topics discussed in this proceeding.
This research was partially supported by the Deutsche
Forschungsgemeinschaft (DFG, German Research Foundation) under grant
396021762-TRR 257.


\bibliographystyle{JHEP}
\bibliography{proceeding_radcor.bib}

\end{document}